\documentclass[orivec,runningheads]{llncs}
\usepackage[utf8]{inputenc}
\usepackage[T1]{fontenc}
\usepackage{amsmath}
\usepackage{longtable}
\usepackage{booktabs}
\usepackage{siunitx}
\usepackage{makecell}
\usepackage{caption}
\usepackage{pifont}
\usepackage{enumitem}
\usepackage{tikz}\usetikzlibrary{shapes,arrows,backgrounds,calc,automata}
\usepackage{nicefrac}
\usepackage{hyperref}
\usepackage{fancyvrb}
\usepackage{graphicx}

\begin{document}

\title{Generalised Arc Consistency via the Synchronised Product of Finite Automata wrt a Constraint}
\titlerunning{GAC via the Synchronised Product of Finite Automata}
\author{Nicolas~Beldiceanu}
\institute{IMT Atlantique, LS2N, UMR CNRS 6004, F-44307 Nantes, France\\
\email{nicolas.beldiceanu@imt-atlantique.fr}}
\maketitle

\begin{abstract}
Given an $m$ by $n$ matrix $V$ of domain variables $v_{i,j}$ (with $i$ from $1$ to $m$ and $j$ from $1$ to $n$), where each row $i$ must be accepted by a specified Deterministic Finite Automaton (DFA) $\mathcal{A}_i$ and each column $j$ must satisfy the same constraint $\texttt{ctr}$, we show how to use the \emph{synchronised product of DFAs wrt constraint} $\texttt{ctr}$ to obtain a Berge-acyclic decomposition ensuring  Generalised Arc Consistency (GAC). Such decomposition consists of one \texttt{regular} and $n$ \texttt{table} constraints.
We illustrate the effectiveness of this method by solving a hydrogen distribution problem, finding optimal solutions and proving optimality quickly.
\end{abstract}

\section{Introduction}

The synchronised product of automata is a key technique, well established in formal methods since the 1960s~\cite{Hartmanis60}, for modelling concurrent systems.
However, its potential as an effective propagation mechanism in constraint programming for achieving GAC filtering~\cite{HandbookCP2006} for specific matrix models was not investigated.

\paragraph{Motivation}
Motivated by a problem of generating cyclic schedules for continuous hydrogen supply to customers via containers,\footnote{Companies producing and distributing green hydrogen like Lhyfe, see~\url{https://www.lhyfe.com/}, were the source of our hydrogen distribution problem.} we exploit the concept of synchronised product of automata wrt a constraint to obtain GAC on a subset of the constraints, specifically relaxing container capacities and customer demands.
In this context, the matrix $V$'s elements, $v_{i,j}$ (with $i\in[1,m]$, $m=$ number of containers; $j\in[1,n]$, $n=$ number of time slots), denote the specific location (refill site or customer) visited by container $i$ at time $j$.

\begin{itemize}[label={\scriptsize\textbullet}]
\item Rows (container $i$'s itinerary) are governed by a \texttt{regular} constraint~\cite{Pesant04}, which enforces a specific cyclic sequence of customer and refill sites.
\item Columns (all containers at time slot $j$) are restricted by an \texttt{alldifferent}~\cite{Regin94}, ensuring no two containers are at the same site simultaneously.
\end{itemize}
To achieve GAC on this conjunction~($C\hspace*{0.7pt}$) of $m$ \texttt{regular} constraints and $n$ \texttt{alldif\-ferent} constraints, we propose a technique based on the synchronised product construction.
This product is formed over the automata corresponding to the $m$ rows, and is explicitly restricted by the \texttt{alldifferent} constraints of the columns.
The resulting automaton captures the interaction between all row and column constraints, providing a powerful mechanism for filtering out inconsistent values and for directly proving the infeasibility of conjunction~($C\hspace*{0.7pt}$) even before search: the right automaton-based abstraction eliminates backtracking.

\paragraph{Contributions} Our contributions include:
\begin{enumerate}
\item For a matrix model where each row must satisfy a given DFA and each column fulfils a same given constraint, a method that captures such a conjunction~($C\hspace*{0.7pt}$) by generating a single synchronised product automaton.
\item The demonstration that the resulting synchronised product automaton can be reformulated as a Berge-acyclic~\cite{BeeriFaginMaierYannakakis83} conjunction of one \texttt{regular}~\cite{Pesant04} and $n$ \texttt{table}~\cite{LhommeR05} constraints, which permits obtaining GAC.
\item To address the original problem by considering the containers’ capacities and customers’ demands, we extract minimal size solutions from the synchronised product automaton and solve a tiny MIP model for each extracted solution.
\end{enumerate}

\paragraph{Paper organisation}
(A)~We begin with the \emph{necessary theoretical framework}.
In Sect.~\ref{sec:background} we recall the notion of a synchronised product of automata wrt a constraint. Building upon this, in Sect.~\ref{sec:reformulation} we give a Berge-acyclic reformulation of a synchronised automata as a conjunction of a \texttt{regular} and \texttt{table} constraints.
(B)~In Sect.~\ref{sec:problem_description} we \emph{introduce our motivating application}, the Hydrogen Distribution Problem (HDP).
(C)~Finally, we \emph{present the integrated solution}.
In Sect.~\ref{sec:cyclic_disjunctive_constraints} we demonstrate the core contribution, showing how the reformulated synchronised product automaton solves the cyclic and disjunctive aspects of the HDP without backtracking. This leads to Sect.~\ref{sec:capacity_constraints}, where we detail a tiny MIP model to handle the remaining capacity and demand constraints.

\section{Background}\label{sec:background}

In this section, we recall the necessary definitions for the rest of the article.
\begin{definition}[synchronised product of automata]\label{def:synchronised_prod}
Given $m$ DFA $\mathcal{A}_i=(Q_i,\Sigma_i,\delta_i,q_{0i},F_i)$ (with $i\in[1,m]$), where each automaton $\mathcal{A}_i$ is defined by
its set of states $Q_i$, its local input alphabet $\Sigma_i$, its transition function $\delta_i$, its initial state $q_{0i}$, and its set of final states $F_i$,
the \emph{synchronised product} $\mathcal{A}_1\times \mathcal{A}_2 \times \dots \times \mathcal{A}_m$, denoted $\mathcal{A}_{\mathit{prod}}=(Q_{\mathit{prod}},
\Sigma_{\mathit{prod}},
\delta_{\mathit{prod}},
q_{0\mathit{prod}},
F_{\mathit{prod}})$, is defined as:
\begin{itemize}[label={\scriptsize\textbullet}]
\item $Q_{\mathit{prod}}=Q_1 \times Q_2 \times \dots \times Q_m$,                                  \hspace*{1.5cm}{\scriptsize\textbullet}\hspace*{0.18cm}$q_{0\mathit{prod}}=(q_{0,1},q_{0,2},\dots,q_{0,m})$,
\item $\Sigma_{\mathit{prod}}=\Sigma_1 \times \Sigma_2 \times \dots \times \Sigma_m$, \hspace*{1.5cm}{\scriptsize\textbullet}\hspace*{0.18cm}$F_{\mathit{prod}}=F_1\times F_2\times\dots\times F_m$,
\item $\delta_{\mathit{prod}}((q_1,q_2,\dots,q_m),(\ell_1,\ell_2,\dots,\ell_m))\!=\!(\delta_1(q_1,\ell_1),\delta_2(q_2,\ell_2)),\dots,\delta_m(q_m,\ell_m))$.
\end{itemize}
\end{definition}

The definition of a \emph{synchronised product of automata wrt a constraint} extends Definition~\ref{def:synchronised_prod}
by introducing a specific synchronisation constraint on the product of the local alphabets $\Sigma_1,\Sigma_2,\dots,\Sigma_m$ of automata $\mathcal{A}_1,\mathcal{A}_2,\dots,\mathcal{A}_m$.

\begin{definition}[synchronised product of automata wrt a constraint]\label{def:synchronised_prod_ctr}
Given $m$ DFA $\mathcal{A}_i=(Q_i,\Sigma_i,\delta_i,q_{0i},F_i)$ (with $i\in[1,m]$) and a constraint $\mathcal{C}$ on $m$ variables,
the \emph{synchronised product of automata $\mathcal{A}_1,\mathcal{A}_2,\dots,\mathcal{A}_m$ wrt $\mathcal{C}$}, denoted $\mathcal{A}_{\mathit{prod}}^{\mathcal{C}}$, is an automaton induced by $\mathcal{A}_{\mathit{prod}}$ on the same alphabet $\Sigma_{\mathit{prod}}$, where the transition function is restricted to $\delta_{\mathit{prod}}^{\mathcal{C}}((q_1,q_2,\dots,q_m),(\ell_1,\ell_2,\dots,\ell_m))=(\delta_1(q_1,\ell_1),$ $\delta_2(q_2,\ell_2)),\dots,$ $\delta_m(q_m,\ell_m))$ where constraint $\mathcal{C}(\ell_1,\ell_2,\dots,\ell_m)$ holds.
\end{definition}
\textbf{Notation:} $\min(\mathcal{A}_{\mathit{prod}}^{\mathcal{C}})$ denotes the minimal automaton associated with $\mathcal{A}_{\mathit{prod}}^{\mathcal{C}}$.
\textbf{Remark~1:} To efficiently generate transitions exiting a state $(q_1,q_2,\dots,q_m)$ of $\mathcal{A}_{\mathit{prod}}^{\mathcal{C}}$, we use a CP model that imposes constraint $\mathcal{C}$ on the labels of the outgoing transitions from states $q_1,q_2,\dots,q_m$ of $\mathcal{A}_1, \mathcal{A}_2,\dots,\mathcal{A}_m$.

We recall the definition of equivalent letters in a DFA.
The equivalence between two letters means that the DFA does not distinguish them in terms of state change, regardless of their position.
This will be used to reduce the size of the synchronised product's alphabet.
\begin{definition}[equivalent letters]\label{def:equivalent_letters}
Given a deterministic finite automata $\mathcal{A}=(Q,\Sigma,\delta,q_0,F)$, \emph{two distinct letters $\ell_1$ and $\ell_2$ of $\Sigma$ are equivalent}, if and only if $\forall q\in Q, \delta(q,\ell_1)=\delta(q,\ell_2)$.
\end{definition}

\section{Berge-Acyclic Reformulation via Synchronised Product}\label{sec:reformulation}

The challenge lies in propagating GAC over the conjunction of the $m$ row-wise \texttt{regular} constraints and the $n$ column-wise constraints $\mathcal{C}$.
By constructing the synchronised product automaton $\mathcal{A}_{\mathit{prod}}^{\mathcal{C}}$, we obtain a Berge-acyclic~\cite{BeeriFaginMaierYannakakis83} reformulation.
This property guarantees that achieving GAC on the resulting sub-constraints is equivalent to achieving GAC on the original system.
We reformulate the original matrix model
\begin{equation}
\bigwedge_{i\in[1,m]}\texttt{regular}(\mathcal{A}_i,\langle v_{i,1},v_{i,2},\dots,v_{i,n}\rangle)~~\land~\bigwedge_{j\in[1,n]}\mathcal{C}(v_{1,j},v_{2,j},\dots,v_{m,j})
\end{equation}
into the following conjunction of one \texttt{regular} and $n$ \texttt{table} constraints:
\begin{equation}\label{eq:reformulation}
\begin{split}
\texttt{regular}(\overline{\min(\mathcal{A}_{\mathit{prod}}^{\mathcal{C}})},\langle w_1,w_2,\dots,w_n\rangle)~~~\land\hspace*{55pt} \\
\bigwedge_{j\in[1,n]}\texttt{table}\left(
\left\langle\begin{array}{ccccc}
v_{1,j} & v_{2,j} & \dots & v_{m,j} & w_j
\end{array}\right\rangle,
\left\langle\begin{array}{cccc|c}
\ell_{1,1} & \ell_{2,1} & \dots & \ell_{m,1} & \ell_{m+1,1} \\
\ell_{1,2} & \ell_{2,2} & \dots & \ell_{m,2} & \ell_{m+1,2} \\
\vdots & \vdots &  \ddots & \vdots & \vdots \\
\ell_{1,p} & \ell_{2,p} & \dots & \ell_{m,p} & \ell_{m+1,p}
\end{array}\right\rangle,
\right)
\end{split}
\end{equation}
\begin{enumerate}[label=\Roman*]
\item The \texttt{regular} constraint, i.e. the first term of (\ref{eq:reformulation}):
	\begin{itemize}[label={\scriptsize\textbullet}]
	\item
	$\overline{\min(\mathcal{A}_{\mathit{prod}}^{\mathcal{C}})}$ is the minimal synchronised product automaton,
	$\min(\mathcal{A}_{\mathit{prod}}^{\mathcal{C}})$, whose input alphabet is a reduced set of unique integer labels.
	To reduce the size of the input alphabet of $\overline{\min(\mathcal{A}_{\mathit{prod}}^{\mathcal{C}})}$,
	each label represents an equivalence class of one or more feasible transition tuples from $\min(\mathcal{A}_{\mathit{prod}}^{\mathcal{C}})$ (as defined in Def.~\ref{def:equivalent_letters}).
	These labels are assigned to the auxiliary variables $w_1, w_2, \dots, w_n$ as specified by the \texttt{table} constraints in~\ref{item:2}.	
	\item
	The auxiliary sequence of variables $w_1, w_2, \dots, w_n$ represents the sequence of global column labels read by the automaton $\overline{\min(\mathcal{A}_{\mathit{prod}}^{\mathcal{C}})}$.
	\end{itemize}
\item\label{item:2} The \texttt{table} constraints, i.e. the second term of (\ref{eq:reformulation}):
	\begin{itemize}[label={\scriptsize\textbullet}]
	\item Each \texttt{table} constraint functionally determines the auxiliary variable $w_j$ from the original column variables $v_{1,j}, v_{2,j}, \dots, v_{m,j}$.
	\item The table itself lists all feasible tuples where each row, $\ell_{1,k}, \ell_{2,k}, \dots, \ell_{m,k},$ $\ell_{m+1,k}$, consists of:
		\begin{itemize}[label={\scriptsize\textbullet}]
		\item A valid assignment of values to the original column variables, that is, one of the feasible tuples associated with the transitions
		        of the minimal synchronised product automaton $\min(\mathcal{A}_{\mathit{prod}}^{\mathcal{C}})$.
		\item The corresponding unique global label $\ell_{m+1,k}$ from the reduced global alphabet that the automaton $\overline{\min(\mathcal{A}_{\mathit{prod}}^{\mathcal{C}})}$ reads for this column.
		\end{itemize}
	\end{itemize}
\end{enumerate}
\textbf{Remark~2:}
Once the original problem’s variables are fixed, variables $w_1, \dots, w_n$ will get fixed by the \texttt{table} constraints.
Thus, there is no need to expose $w_1, \dots, w_n$, and search can be performed solely by enumerating potential values for the original variables of matrix $V$.

\noindent\textbf{Remark~3:}
The constraint network associated with the reformulation given by~(\ref{eq:reformulation}) is Berge-acyclic~\cite{BeeriFaginMaierYannakakis83}, since (\emph{i})~there is at most one shared variable between any two constraints, and (\emph{ii})~the hypergraph of the constraint network contains no cycles.

\paragraph{Limits of the Approach}
The primary limitation lies in the potential state-space complexity of the resulting automaton, $\min(\mathcal{A}_{\mathit{prod}}^{\mathcal{C}})$.
In the worst case, the number of states, $|Q_{\mathit{prod}}^{\mathcal{C}}|$, is bounded by the product $\prod_{i=1}^m |Q_i|$,
and the size of the input alphabet is bounded by the number of feasible assignments to $\mathcal{C}$.
Crucially, the size of $\min(\mathcal{A}_{\mathit{prod}}^{\mathcal{C}})$ is independent of $n$ (the number of columns).
Since the complexity only depends on the number of rows ($m$) and the tightness of constraint $\mathcal{C}$, the approach is well suited for our motivating problem, where the number of containers ($m$) is limited, but the time horizon ($n$) is large.
Furthermore, the size of the automaton is often mitigated by two factors:
\begin{itemize}[label={\scriptsize\textbullet}]
\item
Constraint Pruning and State Reachability:
Both a tighter column constraint $\mathcal{C}$ and more restrictive row automata $\mathcal{A}_i$ prune transition possibilities,
making some states in $\mathcal{A}_{\mathit{prod}}$ unreachable, leading to a smaller $\min(\mathcal{A}_{\mathit{prod}}^{\mathcal{C}})$.
\item
Alphabet Minimisation: If the automata associated with the matrix’s rows contain equivalent letters, the alphabet substitution process may significantly reduce the size of the input alphabet of the automaton $\overline{\min(\mathcal{A}_{\mathit{prod}}^{\mathcal{C}})}$.
\end{itemize}

\section{The Hydrogen Distribution Problem (HDP)}\label{sec:problem_description}

We consider a set of locations, $1,2,\dots,\mathit{loc}$, where `$1$' denotes the production site where each container must be fully refilled for a minimum duration of $R$ time units. The locations $2, 3, \dots, \mathit{loc}$ denote the customer sites, each with a constant instantaneous demand $D_k$, with $k$ in $[2,\mathit{loc}]$.
We also consider a set of $m$ containers, $1,2,\dots,m$, with individual capacities $C_i$, $i$ ranging from $1$ to $m$. 

\paragraph{Continuous Service and Scheduling Constraints}
Since all customers require uninterrupted supply, a strict, instantaneous container swap is enforced: as soon as one container is removed from any site (customer or production), a new container must immediately take its place. Consequently, exactly one container is present at each location at any given moment, meaning $m=\mathit{loc}$. The following constraints apply:
\begin{itemize}[label={\scriptsize\textbullet}]
\item\hspace*{0pt}[\emph{Cyclic Constraint}]
Each container follows a predetermined cyclic sequence of stops (customers and production sites), where the starting location is arbitrary. Inter-site travel time is considered negligible, as the original problem was broken down into subproblems that cluster together nearby locations.
\item\hspace*{0pt}[\emph{Disjunctive Constraint}]
At any instant, each customer site $k \in [2, \mathit{loc}]$ must be served by one container.
\item\hspace*{0pt}[\emph{Demand and Capacity Constraints}]
At any instant, each customer site $k \in [2, \mathit{loc}]$ must be served by a non-empty container that meets its demand $D_k$.
\end{itemize}
\paragraph{Objective}\hspace*{-3pt}:
construct a cyclic schedule that maximises the cycle duration, meaning that it minimises the number of reloads required when assuming an infinite time frame in which the cyclic schedule is repeated infinitely.

\section{Managing Cyclic\hspace*{-0.5pt} \&\hspace*{-0.5pt} Disjunctive Constraints in the HDP}\label{sec:cyclic_disjunctive_constraints}

To handle the cyclic and disjunctive constraints, we generate a synchronised product of automata with respect to a constraint: (\emph{i})~the cyclic restriction of each container is encoded as a regular expression that is converted into a finite automaton; (\emph{ii})~the disjunctive constraint, ensuring that each customer is served by only one container at a time, is modelled using an \texttt{alldifferent} constraint.
This synchronised product automaton wrt \texttt{alldifferent} accepts certain words.
A word specifies, for each container, the precise order of visited sites that adheres to both the cyclic and disjunctive requirements. These words represent the layout of feasible solutions for the HDP when relaxing the demand and capacity constraints.

We present two examples of synchronised automata that were generated from our HDP.
Then, we provide statistics on the sizes of the automata that were generated using this approach on a test set of 118 instances.

\begin{example}[infeasible system of cyclic constraints]\label{ex:infeasible}
Consider the four expressions $R_1$, $R_2$, $R_3$ and $R_4$, which respectively restrict the four rows (container itineraries) of a matrix $V$.
The alphabet $\Sigma = \{1, 2, 3, 4\}$ represents the possible locations (refill site/customers).
Rows $R_2, R_3$, and $R_4$ are constrained by a cyclic sequence requirement.
A cyclic sequence means that the itinerary must follow a fixed sequence of locations and can start at any point within that sequence.
To break symmetry, $R_1$ is not a cyclic sequence but a fixed sequence.
The constraints are defined as follows:
\begin{itemize}[label={\scriptsize\textbullet}]
\item
$R_1 =$ {\small$2^+ 1^+ 3^+ 1^+ 4^+ 1^+$}, a fixed sequence for symmetry breaking,
\item
$R_ 2 =$ the cyclic sequence associated with $3^+ 4^+ 1^+$. This is formally defined by the regular expression {\small$1^* 3^+ 4^+ 1^+ | 3^* 4^+ 1^+ 3^+ | 4^* 1^+ 3^+ 4^+$},
\item
$R_3 =$ the cyclic sequence associated with $4^+ 1^+ 2^+ 3^+ 1^+$,
\item
$R_4 =$ the cyclic sequence associated with $4^+ 2^+ 1^+$.
\end{itemize}
The row automata associated with $R_1$--$R_4$ enforce these constraints, and
the synchronised product of these automata wrt the \texttt{alldifferent} constraint $\mathcal{C}$ yields an empty automaton.
This shows, without using any search, that there is no solution to this conjunction, regardless of the number of columns in the matrix~$V$.
\end{example}
\begin{figure}[!h]
\begin{tikzpicture}
\begin{scope}[->,>=stealth',shorten >=1pt,auto,node distance=8mm,semithick,
              information text/.style={rounded corners=1pt,inner sep=1ex}]
\node[state,minimum size=4mm,initial,initial where=left,initial distance=0.3cm,initial text=] (a) {\scriptsize a};
\node[state,minimum size=4mm]                                          (b)[above right of=a] {\scriptsize b};
\node[state,minimum size=4mm]                                          (c)[right of=b] {\scriptsize c};
\node[state,minimum size=4mm]                                          (d)[right of=c] {\scriptsize d};
\node[state,minimum size=4mm]                                          (e)[right of=d] {\scriptsize e};
\node[state,minimum size=4mm]                                          (f)[right of=e] {\scriptsize f};
\node[state,minimum size=4mm]                                          (g)[right of=f] {\scriptsize g};
\node[state,minimum size=4mm,accepting]                          (h)[right of=g] {\scriptsize h};
\node[state,minimum size=4mm]                                          (i)[below right of=a] {\scriptsize i};
\node[state,minimum size=4mm]                                          (j)[right of=i] {\scriptsize j};
\node[state,minimum size=4mm]                                          (k)[right of=j] {\scriptsize k};
\node[state,minimum size=4mm]                                          (l)[right of=k] {\scriptsize l};
\node[state,minimum size=4mm]                                          (m)[right of=l] {\scriptsize m};
\node[state,minimum size=4mm]                                          (n)[right of=m] {\scriptsize n};
\node[state,minimum size=4mm,accepting]                          (o)[right of=n] {\scriptsize o};
\path
	(a) edge                      node {\scriptsize \textbf{4}} (b)
	(a) edge                      node[below] {\scriptsize \textbf{5~~~~}} (i)
	(b) edge                      node {\scriptsize \textbf{1}} (c)
	(b) edge [loop above] node {\scriptsize \textbf{4}} (b)
	(c) edge [loop above] node {\scriptsize \textbf{1}} (c)
	(c) edge                      node {\scriptsize \textbf{7}} (d)
	(d) edge                      node {\scriptsize \textbf{2}} (e)
	(d) edge [loop above] node {\scriptsize \textbf{7}} (d)
	(e) edge [loop above] node {\scriptsize \textbf{2}} (e)
	(e) edge                      node {\scriptsize \textbf{10}} (f)
	(f) edge                      node {\scriptsize \textbf{9}} (g)
	(f) edge [loop above] node {\scriptsize \textbf{10}} (f)
	(g) edge                      node {\scriptsize \textbf{3}} (h)
	(g) edge [loop above] node {\scriptsize \textbf{9}} (g)
	(h) edge [loop above] node {\scriptsize \textbf{3}} (h)
	(i) edge                     node[below] {\scriptsize \textbf{2}} (j)
	(i) edge [loop below] node[below] {\scriptsize \textbf{5}} (i)
	(j) edge [loop below] node[below] {\scriptsize \textbf{2}} (j)
	(j) edge                     node[below] {\scriptsize \textbf{6}} (k)
	(k) edge [loop below] node[below] {\scriptsize \textbf{6}} (k)
	(k) edge                     node[below] {\scriptsize \textbf{8}} (l)
	(l) edge                   node[below] {\scriptsize \textbf{3}} (m)
	(l) edge [loop below] node[below] {\scriptsize \textbf{8}} (l)
	(m) edge [loop below] node[below] {\scriptsize \textbf{3}} (m)
	(m) edge                     node[below] {\scriptsize \textbf{9}} (n)
	(n) edge                     node[below] {\scriptsize \textbf{1}} (o)
	(n) edge [loop below] node[below] {\scriptsize \textbf{9}} (n)
	(o) edge [loop below] node[below] {\scriptsize \textbf{1}} (o);
\node at (-0.3cm,-1.6cm) {(A)};
\end{scope}
\begin{scope}[xshift=5.6cm,yshift=0cm,information text/.style={rounded corners,inner sep=1ex}]
\draw node[right,information text] {
\setlength{\tabcolsep}{9pt}
\scriptsize
\begin{tabular}{*{4}{S[table-format=1.0]} | c}
\toprule
\multicolumn{4}{c|}{local column assignment} &
\multicolumn{1}{c}{global letter} \\
$\ell_{1}$ & $\ell_{2}$ & $\ell_{3}$ & $\ell_{4}$ & $\ell_{g}$ \\
\midrule
1 & 2 & 4 & 3 & $\textbf{1}$\\
1 & 3 & 2 & 4 & $\textbf{2}$\\
1 & 4 & 2 & 3 & $\textbf{3}$\\
2 & 1 & 4 & 3 & $\textbf{4}$\\
2 & 3 & 1 & 4 & $\textbf{5}$\\
3 & 1 & 2 & 4 & $\textbf{6}$\\
3 & 2 & 1 & 4 & $\textbf{7}$\\
3 & 4 & 2 & 1 & $\textbf{8}$\\
4 & 1 & 2 & 3 & $\textbf{9}$\\
4 & 3 & 2 & 1 & $\textbf{10}$\\
\bottomrule
\end{tabular}
};
\node at (5.8cm,-1.6cm) {(B)};
\end{scope}
\end{tikzpicture}
\caption{\label{fig:example_cyclic}
(A)~Minimal synchronised product automaton $\overline{\min(\mathcal{A}_{\mathit{prod}}^{\mathcal{C}})}$ with an input alphabet that matches the table’s column $\ell_g$ in Table~(B);
(B)~Tuples of the \texttt{table} constraints used in Reformulation~(\ref{eq:reformulation}).
}
\end{figure}
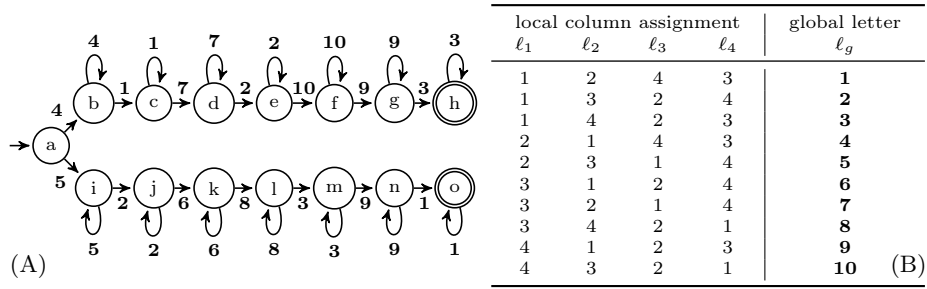

\begin{example}[feasible system of cyclic constraints]\label{ex:feasible}
Now consider a modification of Example~\ref{ex:infeasible} where $R_1$ is kept unchanged,
but $R_2$, $R_3$, and $R_4$ respectively correspond to the cyclic sequences
$4^+ 1^+ 2^+ 3^+ 1^+ $,
$2^+ 4^+ 1^+$, and
$3^+ 4^+ 1^+$.
Part~(A) of Fig.~\ref{fig:example_cyclic} shows the corresponding synchronised product of the automata of $R_1$, $R_2$, $R_3$ and $R_4$ wrt \texttt{alldifferent}.
Part~(B) provides the tuples for the \texttt{table} constraints in Reformulation~(\ref{eq:reformulation}).
This table maps the feasible column assignments $(\ell_1, \ell_2, \ell_3, \ell_4)$, which satisfy the \texttt{alldifferent} constraint and are valid transitions for the row automata's current states, to a unique global letter $\ell_g$, a potential value of the auxiliary variables $w_j$ (with $j\in[1,n]$).
\noindent\textit{Key Observations:}
\begin{itemize}[label={\scriptsize\textbullet}]
\item
The minimal DFAs for $R_1, R_2, R_3, R_4$ have 7, 29, 13, and 13 states.
\item
The synchronised product $\min(\mathcal{A}_{\mathit{prod}}^{\mathcal{C}})$ has only 15 states (a--o), much less than the worst-case product of \num{34307} states.
This reduction is due to the tightness of \texttt{alldifferent}, which makes most combined states unreachable.
\item
Of the $4! = 24$ possible \texttt{alldifferent} assignments for the column variables $\{v_{1,j}, v_{2,j}, v_{3,j}, v_{4,j}\}$, only $10$ are feasible transitions, as shown in Table~(B) of Fig.~\ref{fig:example_cyclic}.
Such pruning of the global alphabet contributes to efficiency.
\end{itemize}
\end{example}

\noindent Table~\ref{table:results} presents observed statistics from 118 test instances of the hydrogen distribution problem, where the number of containers $m$ is fixed per table.
The column constraint $\mathcal{C}$ is \texttt{alldifferent}.
The results in Table~\ref{table:results} confirm that the resulting synchronised product automata are small relative to the theoretical worst-case product of states in our motivating problem.
For $m=4$, the average number of output states is only \num{6.4}, which demonstrates the practical viability of the approach for models with small $m$ and a tight column constraint $\mathcal{C}$.

\begin{table}[!h]
\caption{\normalsize
Observed size of the synchronised automaton ($\min(\mathcal{A}_{\mathit{prod}}^{\mathcal{C}})$) for $m=3$ and $m=4$ containers;
Statistics are given for the minimum ($\min$), maximum ($\max$), sample mean ($\bar{x}$), and sample standard deviation ($s$);
``In'' refers to the properties of the initial DFAs, and ``Out'' refers to the properties of the resulting synchronised product automaton $\overline{\min(\mathcal{A}_{\mathit{prod}}^{\mathcal{C}})}$.
}\label{table:results}
\centering\scriptsize
\setlength{\tabcolsep}{5pt}
\begin{tabular}{rcccc}
\toprule
(82 products with $m=3$ from which 2 were infeasible) & $\min$ & $\max$ & $\bar{x}$ & $s$ \\
\midrule
In states product ($\prod_{i=1}^m |Q_i|$)                                                                        & \num{196} & \num{1805} & \num{437.7} & \num{435.5} \\
Out states (\# states of $\overline{\min(\mathcal{A}_{\mathit{prod}}^{\mathcal{C}})}$)   & 0     & 13     & 5.6     & 2.6     \\
Out alphabet (\# global letters)                                                                                        & 0     & 6       & 3.5     & 1.3     \\
\bottomrule
(179 products with $m=4$ from which 81 were infeasible) & $\min$ & $\max$ & $\bar{x}$ & $s$ \\
\midrule
In states product ($\prod_{i=1}^m |Q_i|$)                                                                        & \num{2058} & \num{229593} & \num{29920.3} & \num{37359.4} \\
Out states (\# states of $\overline{\min(\mathcal{A}_{\mathit{prod}}^{\mathcal{C}})}$)   & 0 & 61 & 6.4 & 9.0 \\
Out alphabet (\# global letters)                                                                                        & 0 & 24 & 4.2 & 4.7 \\
\bottomrule
\end{tabular}
\end{table}

\section{Handling Demand \hspace*{-0.5pt}\&\hspace*{-0.5pt} Capacity Constraints in the HDP}\label{sec:capacity_constraints}

To integrate demand and capacity constraints while maximising the schedule length, we first introduce a property of the synchronised product of automata associated with the HDP to identify a reduced set of essential paths, which we call minimal solutions. We then formulate a tiny MIP problem for each minimal solution to determine the optimal duration of each stage.

\paragraph{Identifying Minimal Solutions}
We present a property of the synchronised product of automata used to model the cyclic and disjunctive constraints of the HDP.

\begin{property}\label{prop:no_cycle}
Given automata $\mathcal{A}_1,\mathcal{A}_2,\dots,\mathcal{A}_m$ that do not contain any cycles involving more than one state, the synchronised product of $\mathcal{A}_1,\mathcal{A}_2,\dots,\mathcal{A}_m$ wrt any constraint does not contain any cycles involving more than one state.
\end{property}

\noindent Property~\ref{prop:no_cycle} can be proved by contradiction.
Assume that the synchronised product automaton contains a cycle with more than one state.
It would then accept an infinite number of words.
This would imply that all automata $\mathcal{A}_i$ (with $i\in[1,m]$) must also contain such cycles, a contradiction.
Since each cyclic constraint of the HDP corresponds to the union of a fixed number of automata with no cycles involving more than one state, this ensures that the synchronised automaton, once self-loops are removed, has a finite number of solutions. They correspond to the possible starting points in the cyclic sequence associated with each container.

\begin{definition}
We call the \emph{minimal set of solutions} of an automaton $\mathcal{A}$ with no cycle involving more than one state, the finite set of solutions of automaton $\mathcal{A}$ from which we discard all self-loops.
\end{definition}

\begin{example}[Continuation of Example~\ref{ex:feasible}]
The synchronised product automaton depicted in Fig.~\ref{fig:example_cyclic} has two minimal solutions: `\textbf{4} \textbf{1} \textbf{7} \textbf{2} \textbf{10} \textbf{9} \textbf{3}' and `\textbf{5} \textbf{2} \textbf{6} \textbf{8} \textbf{3} \textbf{9} \textbf{1}'.
A minimal solution represents a valid sequence of global production stages (locations visited by each container) that satisfies the cyclic and disjunctive constraints.
The first minimal solution corresponds to the sequences visited by containers 1 to 4: `2 1 3 1 4 4 1', `1 2 2 3 3 1 4', `4 4 1 2 2 2 2', and `3 3 4 4 1 3 3'.
The first location in each sequence is the starting location for that container.
\end{example}

\paragraph{Formulating a Compact MIP Model}
To handle the demand and capacity constraints, we create a tiny MIP model for each minimal solution identified in the first step.
The MIP determines the optimal duration of each stage in the minimal solution, ensuring demand fulfilment and capacity limits are respected, while maximising the total duration.
Given a minimal solution of length $n$, we create $n$ integer variables $p_1,p_2,\dots,p_n$, where $p_k$ (with $k\in[1,n]$) is the duration of the $k$-th stage.
We have the following constraints:

\noindent \textbullet~[Refill]
At each stage $k$ (with $k\in[1,n]$), one of the containers is always at the production site for refill/loading. Thus, the minimum duration of any stage $p_k$ is set to the refill time $R$, i.e.~ $p_k \geq R$.

\noindent \textbullet~[Demand]
For each container $i \in [1,m]$, we generate a set of constraints associated with the successive locations it visits.
A critical subsequence $\mathcal{S}$ of visited customer locations is defined as the locations visited by container $i$ between two successive visits to the production site. For each such subsequence $\mathcal{S}$, we post a linear capacity constraint: $\sum_{k \in \mathcal{S}} D_k \cdot p_k \leq C_i$.
$C_i$ is the capacity of container $i$, and
$D_k$ is the demand of the customer location being visited at stage $k$.

\noindent \textbullet~[Objective Function]
The objective is to maximise the overall schedule length, which is the sum of the durations of all stages: $\text{Maximise} \sum_{k=1}^{n} p_k$.

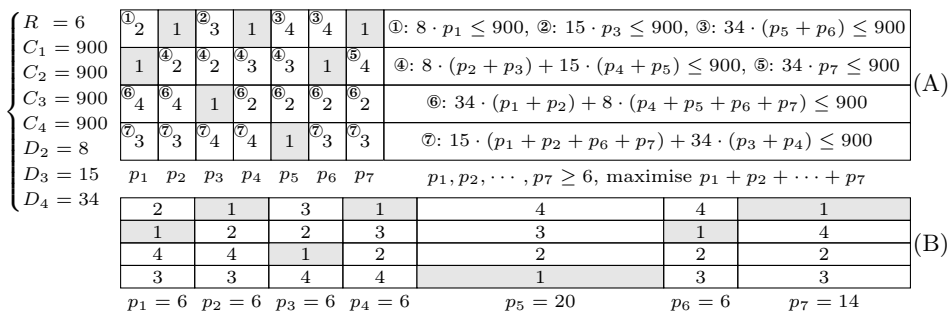
\begin{figure}[!h]
\center
\begin{tikzpicture}
\begin{scope}[scale=0.5]
\draw[draw=white] (0,-3.9) rectangle (1,-3) node[pos=.5] {\scriptsize $p_1$};
\draw[draw=white] (1,-3.9) rectangle (2,-3) node[pos=.5] {\scriptsize $p_2$};
\draw[draw=white] (2,-3.9) rectangle (3,-3) node[pos=.5] {\scriptsize $p_3$};
\draw[draw=white] (3,-3.9) rectangle (4,-3) node[pos=.5] {\scriptsize $p_4$};
\draw[draw=white] (4,-3.9) rectangle (5,-3) node[pos=.5] {\scriptsize $p_5$};
\draw[draw=white] (5,-3.9) rectangle (6,-3) node[pos=.5] {\scriptsize $p_6$};
\draw[draw=white] (6,-3.9) rectangle (7,-3) node[pos=.5] {\scriptsize $p_7$};
\draw (0, 0) rectangle (1, 1) node[pos=.5] {\scriptsize 2};
\draw[fill=gray!20] (0,-1) rectangle (1, 0) node[pos=.5] {\scriptsize 1};
\draw (0,-2) rectangle (1,-1) node[pos=.5] {\scriptsize 4};
\draw (0,-3) rectangle (1,-2) node[pos=.5] {\scriptsize 3};
\draw[fill=gray!20] (1, 0) rectangle (2, 1) node[pos=.5] {\scriptsize 1};
\draw (1,-1) rectangle (2, 0) node[pos=.5] {\scriptsize 2};
\draw (1,-2) rectangle (2,-1) node[pos=.5] {\scriptsize 4};
\draw (1,-3) rectangle (2,-2) node[pos=.5] {\scriptsize 3};
\draw (2, 0) rectangle (3, 1) node[pos=.5] {\scriptsize 3};
\draw (2,-1) rectangle (3, 0) node[pos=.5] {\scriptsize 2};
\draw[fill=gray!20] (2,-2) rectangle (3,-1) node[pos=.5] {\scriptsize 1};
\draw (2,-3) rectangle (3,-2) node[pos=.5] {\scriptsize 4};
\draw[fill=gray!20] (3, 0) rectangle (4, 1) node[pos=.5] {\scriptsize 1};
\draw (3,-1) rectangle (4, 0) node[pos=.5] {\scriptsize 3};
\draw (3,-2) rectangle (4,-1) node[pos=.5] {\scriptsize 2};
\draw (3,-3) rectangle (4,-2) node[pos=.5] {\scriptsize 4};
\draw (4, 0) rectangle (5, 1) node[pos=.5] {\scriptsize 4};
\draw (4,-1) rectangle (5, 0) node[pos=.5] {\scriptsize 3};
\draw (4,-2) rectangle (5,-1) node[pos=.5] {\scriptsize 2};
\draw[fill=gray!20] (4,-3) rectangle (5,-2) node[pos=.5] {\scriptsize 1};
\draw (5, 0) rectangle (6, 1) node[pos=.5] {\scriptsize 4};
\draw[fill=gray!20] (5,-1) rectangle (6, 0) node[pos=.5] {\scriptsize 1};
\draw (5,-2) rectangle (6,-1) node[pos=.5] {\scriptsize 2};
\draw (5,-3) rectangle (6,-2) node[pos=.5] {\scriptsize 3};
\draw[fill=gray!20] (6, 0) rectangle (7, 1) node[pos=.5] {\scriptsize 1};
\draw (6,-1) rectangle (7, 0) node[pos=.5] {\scriptsize 4};
\draw (6,-2) rectangle (7,-1) node[pos=.5] {\scriptsize 2};
\draw (6,-3) rectangle (7,-2) node[pos=.5] {\scriptsize 3};
\draw[draw=white] (7,-3.9) rectangle (21,-3) node[pos=.5] {\scriptsize $p_1,p_2,\cdots,p_7\geq 6$, maximise $p_1+p_2+\cdots+p_7$};
\draw (7,0) rectangle (21,1) node[pos=.5] {\scriptsize \ding{172}: $8\cdot p_1\leq 900$, \ding{173}: $15\cdot p_3\leq 900$, \ding{174}: $34\cdot(p_5+p_6)\leq 900$};
\draw (7,-1) rectangle (21,0) node[pos=.5] {\scriptsize \ding{175}: $8\cdot(p_2+p_3)+15\cdot(p_4+p_5)\leq 900$, \ding{176}: $34\cdot p_7 \leq 900$};
\draw (7,-2) rectangle (21,-1) node[pos=.5] {\scriptsize \ding{177}: $34\cdot(p_1+p_2)+8\cdot(p_4+p_5+p_6+p_7)\leq 900$};
\draw (7,-3) rectangle (21,-2) node[pos=.5] {\scriptsize \ding{178}: $15\cdot(p_1+p_2+p_6+p_7)+34\cdot(p_3+p_4)\leq 900$};
\node at (0.2cm,0.8cm) {\scriptsize\ding{172}};
\node at (2.2cm,0.8cm) {\scriptsize\ding{173}};
\node at (4.2cm,0.8cm) {\scriptsize\ding{174}};
\node at (5.2cm,0.8cm) {\scriptsize\ding{174}};
\node at (1.2cm,-0.2cm) {\scriptsize\ding{175}};
\node at (2.2cm,-0.2cm) {\scriptsize\ding{175}};
\node at (3.2cm,-0.2cm) {\scriptsize\ding{175}};
\node at (4.2cm,-0.2cm) {\scriptsize\ding{175}};
\node at (6.2cm,-0.2cm) {\scriptsize\ding{176}};
\node at (0.2cm,-1.2cm) {\scriptsize\ding{177}};
\node at (1.2cm,-1.2cm) {\scriptsize\ding{177}};
\node at (3.2cm,-1.2cm) {\scriptsize\ding{177}};
\node at (4.2cm,-1.2cm) {\scriptsize\ding{177}};
\node at (5.2cm,-1.2cm) {\scriptsize\ding{177}};
\node at (6.2cm,-1.2cm) {\scriptsize\ding{177}};
\node at (0.2cm,-2.2cm) {\scriptsize\ding{178}};
\node at (1.2cm,-2.2cm) {\scriptsize\ding{178}};
\node at (2.2cm,-2.2cm) {\scriptsize\ding{178}};
\node at (3.2cm,-2.2cm) {\scriptsize\ding{178}};
\node at (5.2cm,-2.2cm) {\scriptsize\ding{178}};
\node at (6.2cm,-2.2cm) {\scriptsize\ding{178}};
\node at (21.55cm,-1cm) {(A)};
\end{scope}
\begin{scope}[yshift=-2.3cm,scale=0.5]								
\draw[draw=white]   (0,-2.5) rectangle (1.97,-1.9) node[pos=.5] {\scriptsize $p_1=6$};
\draw               (0, 0  ) rectangle (1.97, 0.6) node[pos=.5] {\scriptsize 2};
\draw[fill=gray!20] (0,-0.6) rectangle (1.97, 0  ) node[pos=.5] {\scriptsize 1};
\draw               (0,-1.2) rectangle (1.97,-0.6) node[pos=.5] {\scriptsize 4};
\draw               (0,-1.8) rectangle (1.97,-1.2) node[pos=.5] {\scriptsize 3};

\draw[draw=white]   (1.97,-2.5) rectangle (3.94,-1.9) node[pos=.5] {\scriptsize $p_2=6$};
\draw[fill=gray!20] (1.97, 0  ) rectangle (3.94, 0.6) node[pos=.5] {\scriptsize 1}; 
\draw               (1.97,-0.6) rectangle (3.94, 0  ) node[pos=.5] {\scriptsize 2};
\draw               (1.97,-1.2) rectangle (3.94,-0.6) node[pos=.5] {\scriptsize 4};
\draw               (1.97,-1.8) rectangle (3.94,-1.2) node[pos=.5] {\scriptsize 3};

\draw[draw=white]   (3.94,-2.5) rectangle (5.91,-1.9) node[pos=.5] {\scriptsize $p_3=6$};
\draw               (3.94, 0  ) rectangle (5.91, 0.6) node[pos=.5] {\scriptsize 3}; 
\draw               (3.94,-0.6) rectangle (5.91, 0  ) node[pos=.5] {\scriptsize 2};
\draw[fill=gray!20] (3.94,-1.2) rectangle (5.91,-0.6) node[pos=.5] {\scriptsize 1};
\draw               (3.94,-1.8) rectangle (5.91,-1.2) node[pos=.5] {\scriptsize 4};

\draw[draw=white]   (5.91,-2.5) rectangle (7.88,-1.9) node[pos=.5] {\scriptsize $p_4=6$};
\draw[fill=gray!20] (5.91, 0  ) rectangle (7.88, 0.6) node[pos=.5] {\scriptsize 1}; 
\draw               (5.91,-0.6) rectangle (7.88, 0  ) node[pos=.5] {\scriptsize 3};
\draw               (5.91,-1.2) rectangle (7.88,-0.6) node[pos=.5] {\scriptsize 2};
\draw               (5.91,-1.8) rectangle (7.88,-1.2) node[pos=.5] {\scriptsize 4};

\draw[draw=white]   (7.88,-2.5) rectangle (14.44,-1.9) node[pos=.5] {\scriptsize $p_5=20$};
\draw               (7.88, 0  ) rectangle (14.44, 0.6) node[pos=.5] {\scriptsize 4}; 
\draw               (7.88,-0.6) rectangle (14.44, 0  ) node[pos=.5] {\scriptsize 3};
\draw               (7.88,-1.2) rectangle (14.44,-0.6) node[pos=.5] {\scriptsize 2};
\draw[fill=gray!20] (7.88,-1.8) rectangle (14.44,-1.2) node[pos=.5] {\scriptsize 1};

\draw[draw=white]   (14.44,-2.5) rectangle (16.41,-1.9) node[pos=.5] {\scriptsize $p_6=6$};
\draw               (14.44, 0  ) rectangle (16.41, 0.6) node[pos=.5] {\scriptsize 4}; 
\draw[fill=gray!20] (14.44,-0.6) rectangle (16.41, 0  ) node[pos=.5] {\scriptsize 1};
\draw               (14.44,-1.2) rectangle (16.41,-0.6) node[pos=.5] {\scriptsize 2};
\draw               (14.44,-1.8) rectangle (16.41,-1.2) node[pos=.5] {\scriptsize 3};

\draw[draw=white]   (16.41,-2.5) rectangle (21,-1.9) node[pos=.5] {\scriptsize $p_7=14$};
\draw[fill=gray!20] (16.41, 0  ) rectangle (21, 0.6) node[pos=.5] {\scriptsize 1}; 
\draw               (16.41,-0.6) rectangle (21, 0  ) node[pos=.5] {\scriptsize 4};
\draw               (16.41,-1.2) rectangle (21,-0.6) node[pos=.5] {\scriptsize 2};
\draw               (16.41,-1.8) rectangle (21,-1.2) node[pos=.5] {\scriptsize 3};
\node at (21.55cm,-0.66cm) {(B)};
\end{scope}
\begin{scope}[xshift=-0.1cm,yshift=-0.85cm,scale=0.5]
\node at (-1.2cm,0cm) {\scriptsize
$
  \begin{cases}
    R   \hspace*{3pt}= 6 \\
    C_1 = 900 \\
    C_2 = 900 \\
    C_3 = 900 \\
    C_4 = 900 \\
    D_2 = 8 \\
    D_3 = 15 \\
    D_4 = 34
  \end{cases}
$
};
\end{scope}
\end{tikzpicture}
\caption{\label{fig:example_capacity}
(A)~MIP model stated for the first minimal solution of the synchronised product automaton of Fig.~2, and (B) the corresponding optimal solution that maximises the total duration. Grey areas indicate the refill site.}
\end{figure}

\begin{example}
Fig.~\ref{fig:example_capacity} illustrates the linear constraints that are created for the first minimal solution of `\textbf{4} \textbf{1} \textbf{7} \textbf{2} \textbf{10} \textbf{9} \textbf{3}' of the automaton depicted in Fig.~\ref{fig:example_cyclic}.
\end{example}

\noindent The total time it took to solve the 118 instances of our test case, including generating cyclic automata and their synchronised product, searching for optimal solutions, and proving optimality, was 5.2s on an M2 Ultra using SICStus 4.8.0. For reproducibility purpose annotated code and instances are in the appendix.

\section{Conclusion}\label{sec:conclusion}

We used a well-known construction from concurrent system modelling to obtain a reformulation performing GAC on a conjunction of constraints on a matrix of variables, where each row verifies a \texttt{regular} constraint and each column verifies the same constraint.
This abstraction immediately rendered our original hydrogen distribution problem obsolete, replacing it with a straightforward optimisation captured by a tiny MIP: ultimately, a good abstraction killed our problem.

If the constraints on the columns differ from column to column, one could extend the approach by using a Multiple Decision Diagram (MDD) corresponding to the unfolding of the synchronised product automaton. This MDD represents the development of the synchronised product automaton, where the constraint of each column is considered at the corresponding level of the MDD.

\section*{Acknowledgments} {\footnotesize Athana\"el Jousselin and Victor Spitzer explained the hydrogen distribution problem to me and created the experimental instances used in this paper. Olivier Peton gave some feedback on an early draft of the paper, and Justin Pearson found the reference for the synchronised product of automata.}

\bibliographystyle{splncs04}
\bibliography{cartesian_prod}

@Incollection{HandbookCP2006,
  author    = {C. Bessi\`ere},
  editor    = {F. Rossi and P. van Beek and T. Walsh},
  title     = {Chapter 3: Constraint Propagation},
  bookTitle = {Handbook of Constraint Programming},
  year      = {2006},
  publisher = {Elsevier},
  pages     = {29--83},
}

@article{Hartmanis60,
  author  = {J. Hartmanis},
  title   = {Symbolic analysis of a decomposition of information processing machines},
  journal = {Information and Control},
  volume  = {3},
  number  = {2},
  pages   = {154--178},
  year    = {1960},
}

@article{BeeriFaginMaierYannakakis83,
  author  = {C. Beeri and R. Fagin and D. Maier and M. Yannakakis},
  title   = {On the Desirability of Acyclic Database Schemes},
  journal = {Journal of the {ACM}},
  volume  = {30},
  number  = {3},
  pages   = {479--513},
  year    = {1983},
}

@inproceedings{LhommeR05,
  author	= {O. Lhomme and {R{\'{e}}gin, J.-C.}},
  title		= {A Fast Arc Consistency Algorithm for n-ary Constraints},
  booktitle	= {20th National Conference on Artificial Intelligence (AAAI-05)},
  pages	= {405--410},
  year	= {2005},
}

@InProceedings{Regin94,
  author        = {{R{\'{e}}gin, J.-C.}},
  title         = {{A} {F}iltering {A}lgorithm for {C}onstraints of {D}ifference in {CSP}},
  booktitle     = {12th National Conference on Artificial Intelligence (AAAI-94)},
  pages         = {362--367},
  year          = {1994},
}

@InProceedings{Pesant04,
  author    = {G. Pesant},
  title     = {{A} {R}egular {L}anguage {M}embership {C}onstraint for {F}inite {S}equences of {V}ariables},
  booktitle = {Principles and Practice of Constraint Programming (CP'2004)},
  year      = {2004},
  editor    = {M. G. Wallace},
  series    = {LNCS},
  volume    = {3258},
  pages     = {482--495},
  publisher = {Springer\nobreakdash-Verlag},
}

\clearpage
\section*{Appendix}

We provide in this appendix supplementary material used for reproducibility purposes: 
\begin{itemize}[label={\scriptsize\textbullet}]
\item
In Section~\ref{sec:instances}, we describe the format of a problem instance and the instances that were used.
\item
In Section~\ref{sec:solutions} we give, for each instance, the optimal solution found and the time spent.
\item
In Section~\ref{sec:code}, we provide instructions to replicate the experiments detailed in the article, along with the associated code.
It also establishes a connection between the paper’s contributions and their corresponding implementation in the code.
\end{itemize}

\subsection{Problem Instances}\label{sec:instances}

Each instance is described by a Prolog fact of the form
{\footnotesize
\begin{verbatim}
    instance(Instance,
             ReloadTime,
             CapacityContainers,
             NeedPerTimeSlotCustomers,
             VisitedCustomersByEachContainer,
             UpperBound).
\end{verbatim}
}
where:
\begin{itemize}[label={\scriptsize\textbullet}]
\item {\footnotesize\texttt{Instance}} is the identifier of the instance.
\item {\footnotesize\texttt{ReloadTime}} is the reload time of all containers at the production site.
\item {\footnotesize\texttt{CapacityContainers}} is a list giving the capacity of each container.
\item {\footnotesize\texttt{NeedPerTimeSlotCustomers}} is a list giving the demand of each customer.
\item {\footnotesize\texttt{VisitedCustomersByEachContainer}} is a list of sublists giving for each customer the cyclic sequence of visited locations. Within a cyclic sequence, a sublist is interpreted as all the permutation of its elements. For example, for the first customer of instance~{\footnotesize\texttt{ia6}}, the argument {\footnotesize\texttt{[[2,3],1]}} is interpreted as the two cyclic sequences {\footnotesize\texttt{[2,3,1]}} and {\footnotesize\texttt{[3,2,1]}}.
\item {\footnotesize\texttt{UpperBound}} is an upper bound of the total duration.
\end{itemize}

\noindent We use the following instances:

\noindent\begin{SaveVerbatim}{VerbInstancesA}
instance(a1,   6, [ 420, 300, 300],      [1,1],     [[2,1,3,1],     [2,1],         [3,1]],                        1000).
instance(a2,   6, [ 300, 300, 300],      [6,4],     [[2,1,3,1],     [2,1,3,1],     [2,3,1]],                      1000).
instance(a3,   6, [ 300, 300, 300],      [6,4],     [[2,1,3,1],     [2,1,3,1],     [3,2,1]],                      1000).
instance(a12,  6, [ 300, 300, 300],      [6,4],     [[2,1,3,1],     [2,1,3,1],     [[2,3],1]],                    1000).
instance(a4,   6, [ 900, 900, 900, 900], [8,15,34], [[2,1,3,1,4,1], [4,1,2,3,1],   [2,4,1],       [3,4,1]],       1000).
instance(a5,   6, [ 900, 900, 900, 900], [8,15,34], [[2,1,3,1,4,1], [4,1,2,3,1],   [2,4,1],       [4,3,1]],       1000).
instance(a6,   6, [ 900, 900, 900, 900], [8,15,34], [[2,1,3,1,4,1], [4,1,2,3,1],   [4,2,1],       [3,4,1]],       1000).
instance(a7,   6, [ 900, 900, 900, 900], [8,15,34], [[2,1,3,1,4,1], [4,1,2,3,1],   [4,2,1],       [4,3,1]],       1000).
instance(a8,   6, [ 900, 900, 900, 900], [8,15,34], [[2,1,3,1,4,1], [4,1,3,2,1],   [2,4,1],       [3,4,1]],       1000).
instance(a9,   6, [ 900, 900, 900, 900], [8,15,34], [[2,1,3,1,4,1], [4,1,3,2,1],   [2,4,1],       [4,3,1]],       1000).
instance(a10,  6, [ 900, 900, 900, 900], [8,15,34], [[2,1,3,1,4,1], [4,1,3,2,1],   [4,2,1],       [3,4,1]],       1000).
instance(a11,  6, [ 900, 900, 900, 900], [8,15,34], [[2,1,3,1,4,1], [4,1,3,2,1],   [4,2,1],       [4,3,1]],       1000).
instance(a13,  6, [ 900, 900, 900, 900], [8,15,34], [[2,1,3,1,4,1], [4,1,[3,2],1], [[4,2],1],     [[4,3],1]],     1000).
instance(ia6,  5, [ 250,1000,1000],      [2,2],     [[[2,3],1],     [2,1],         [3,1]],                         562).
instance(ia7,  5, [ 250, 500,1000],      [3,5],     [[2,1,3,1],     [2,1],         [3,1]],                         300).
instance(ia8,  5, [ 250, 500,1000],      [8,3],     [[2,1,3,1],     [3,1],         [2,1]],                         187).
instance(ib12, 5, [ 250, 500, 500,1000], [1,2,1],   [[2,1,3,1,4,1], [2,1],         [[3,4],1],     [[3,4],1]],      994).
instance(ia13, 5, [ 250, 250,1000],      [5,4],     [[2,1,3,1],     [2,1],         [3,1]],                         250).
instance(ib13, 5, [ 250, 500,1000,1000], [4,4,3],   [[2,1,3,1,4,1], [2,1],         [[3,4],1],     [3,1,[2,4],1]],  505).
instance(ia14, 5, [ 250, 250,1000],      [5,5],     [[2,1,3,1],     [2,1],         [3,1]],                         250).
instance(ib14, 5, [ 250,1000,1000,1000], [3,2,7],   [[2,1,3,1,4,1], [2,1],         [[3,4],1],     [[3,4],1]],      416).
instance(ia21, 5, [ 250, 250,1000],      [2,2],     [[2,1,3,1],     [2,1],         [3,1]],                         625).
instance(ia22, 5, [ 250, 500,1000],      [4,3],     [[2,1,3,1],     [3,1],         [2,1]],                         375).
instance(ic22, 5, [ 500,1000,1000],      [4,6],     [[2,1,3,1],     [2,1],         [3,1]],                         332).
instance(id22, 5, [ 250, 250, 500,1000], [3,1,2],   [[2,1,[3,4],1], [3,1],         [4,1],         [2,1]],          499).
instance(ib23, 5, [ 250,1000,1000,1000], [6,4,2],   [[2,1,3,1,4,1], [3,1],         [[2,4],1],     [2,1,[3,4],1]],  497).
instance(ic23, 5, [ 250, 250, 250,1000], [3,2,2],   [[2,1,3,1,4,1], [3,1],         [[2,4],1],     [[2,4],1]],      493).
instance(ja13, 5, [1000, 250, 250],      [5,4],     [[2,1,3,1],     [2,1],         [3,1]],                         250).
instance(jb13, 5, [1000, 250, 500,1000], [4,4,3],   [[2,1,3,1,4,1], [2,1],         [[3,4],1],     [3,1,[2,4],1]],  505).
instance(jb23, 5, [1000, 250,1000,1000], [6,4,2],   [[2,1,3,1,4,1], [3,1],         [[2,4],1],     [2,1,[3,4],1]],  497).
instance(jc23, 5, [1000, 250, 250, 250], [3,2,2],   [[2,1,3,1,4,1], [3,1],         [[2,4],1],     [[2,4],1]],      493).
instance(kb13, 5, [1000, 250, 500,1000], [4,3,4],   [[2,1,3,1,4,1], [[2,4],1],     [2,1,[3,4],1], [2,1,3,1,4,1]],  630).
instance(kb14b,5, [1000,1000,1000, 500], [8,4,2],   [[2,1,[3,4],1], [3,1,[2,4],1], [4,1,[2,3],1], [2,1]],          432).
instance(kb23b,5, [1000, 500,1000],      [6,8],     [[2,1,3,1],     [2,1],         [3,1]],                         249).
instance(ka23c,5, [1000, 500,1000],      [8,6],     [[2,1,3,1],     [3,1],         [2,1]],                         249).
instance(kb23c,5, [1000, 250,1000],      [3,7],     [[2,1,3,1],     [2,1],         [3,1]],                         284).
instance(kd23c,5, [ 500, 500, 500, 500], [3,3,1],   [[2,1,3,1,4,1], [2,1],         [[3,4],1],     [3,1,[2,4],1]],  497).
instance(kb21d,5, [1000, 500, 500,1000], [2,2,1],   [[2,1,3,1,4,1], [2,1],         [[3,4],1],     [3,1,[2,4],1]], 1166).
instance(kd22d,5, [ 500, 250, 250, 500], [2,1,2],   [[2,1,3,1,4,1], [2,1],         [[3,4],1],     [4,1,[2,3],1]],  583).
instance(la6,  5, [1000, 250,1000],      [2,2],     [[[2,3],1],     [2,1],         [3,1]],                         562).
instance(la7,  5, [1000, 250, 500],      [3,5],     [[2,1,3,1],     [2,1],         [3,1]],                         300).
instance(la8,  5, [1000, 250, 500],      [8,3],     [[2,1,3,1],     [3,1],         [2,1]],                         187).
instance(la12, 5, [ 500,1000,1000],      [3,3],     [[2,1,3,1],     [2,1,3,1],     [2,1,3,1]],                     832).
instance(lb12, 5, [1000, 250, 500, 500], [1,2,1],   [[2,1,3,1,4,1], [2,1],         [[3,4],1],     [[3,4],1]],      994).
instance(la13, 5, [1000, 250, 250],      [4,5],     [[2,1,3,1],     [2,1],         [3,1]],                         250).
instance(lb13, 5, [1000, 250, 500,1000], [4,3,4],   [[2,1,3,1,4,1], [[2,4],1],     [2,1,3,1],     [2,1,3,1,4,1]],  591).
instance(la14, 5, [1000, 250, 250],      [5,5],     [[2,1,3,1],     [2,1],         [3,1]],                         250).
instance(lb14, 5, [1000, 250,1000,1000], [3,2,7],   [[2,1,3,1,4,1], [2,1],         [[3,4],1],     [[3,4],1]],      416).
instance(la21, 5, [ 500, 500, 500],      [1,2],     [[[2,3],1],     [2,1],         [3,1]],                         497).
instance(lb21, 5, [1000, 250, 250],      [2,2],     [[2,1,3,1],     [2,1],         [3,1]],                         625).
instance(ld21, 5, [ 500, 500, 500, 500], [1,1,1],   [[2,1,3,1,4,1], [2,1],         [3,1],         [4,1]],         1000).
instance(la22, 5, [1000, 250, 500],      [4,3],     [[2,1,3,1],     [3,1],         [2,1]],                         375).
instance(lb22, 5, [1000,1000,1000],      [6,6],     [[2,1,3,1],     [2,1],         [3,1]],                         332).
instance(lc22, 5, [1000, 500,1000],      [4,6],     [[2,1,3,1],     [2,1],         [3,1]],                         332).
instance(ld22, 5, [1000, 250, 250, 500], [3,1,2],   [[2,1,[3,4],1], [3,1],         [4,1],         [2,1]],          499).
instance(la23, 5, [1000,1000,1000],      [7,8],     [[2,1,3,1],     [2,1],         [3,1]],                         250).
instance(lb23, 5, [1000, 250, 250,1000], [6,2,2],   [[2,1,[3,4],1], [3,1],         [4,1],         [2,1]],          332).
instance(lc23, 5, [1000, 250, 250, 500], [4,3,2],   [[2,1,3,1,4,1], [3,1],         [4,1],         [2,1]],          375).
instance(la24, 5, [1000, 250, 500, 500], [1,1,1],   [[2,1,3,1,4,1], [4,1],         [2,1],         [3,1]],         1250).
instance(la25, 5, [1000, 250,1000,1000], [1,2,2],   [[2,1,3,1,4,1], [2,1],         [3,1],         [4,1]],         1000).
instance(la26, 5, [1000,1000,1000],      [4,4],     [[2,1,3,1],     [2,1],         [3,1]],                         500).
instance(lb26, 5, [1000, 250,1000],      [5,1],     [[[2,3],1],     [3,1],         [2,1]],                         375).
instance(la30, 5, [1000, 500, 500],      [2,2],     [[2,1,3,1],     [2,1],         [3,1]],                         750).
instance(lb30, 5, [1000,1000,1000],      [3,3],     [[2,1,3,1],     [2,1],         [3,1]],                         666).
instance(le30, 5, [1000, 250, 250, 250], [1,1,1],   [[2,1,3,1,4,1], [2,1],         [3,1],         [4,1]],         1250).
instance(ma6b, 5, [ 500, 250, 500],      [1,1],     [[2,1,3,1],     [3,1],         [2,1]],                         750).
instance(ma7b, 5, [1000, 500, 500],      [2,5],     [[[2,3],1],     [2,1],         [3,1]],                         285).
instance(ma8b, 5, [1000, 500, 500,1000], [1,5,6],   [[3,1,[2,4],1], [2,1],         [3,1],         [4,1]],          300).
instance(ma12b,5, [1000, 500, 500],      [3,3],     [[2,1,3,1],     [2,1],         [3,1]],                         499).
instance(mb12b,5, [ 250, 250, 250],      [1,1],     [[2,1,3,1],     [2,1],         [3,1]],                         500).
instance(mc12b,5, [1000, 500,1000],      [4,3],     [[2,1,3,1],     [3,1],         [2,1]],                         499).
instance(ma13b,5, [ 500, 500, 500],      [3,6],     [[[2,3],1],     [2,1],         [3,1]],                         163).
instance(mb13b,5, [1000, 500, 250, 250], [5,3,5],   [[2,1,3,1,4,1], [4,1],         [[2,3],1],     [[2,3],1]],      294).
instance(ma14b,5, [1000, 500,1000],      [3,10],    [[[2,3],1],     [2,1],         [3,1]],                         192).
instance(mb14b,5, [1000, 500,1000,1000], [8,4,2],   [[2,1,3,1,4,1], [[2,4],1],     [2,1,[3,4],1], [[2,3],1]],      433).
instance(mb21b,5, [1000, 500,1000],      [2,3],     [[2,1,3,1],     [2,1],         [3,1]],                         666).
instance(ma22b,5, [1000, 500, 500],      [5,5],     [[2,1,3,1],     [2,1],         [3,1]],                         300).
instance(mb22b,5, [1000, 250, 250, 250], [2,2,2],   [[2,1,3,1,4,1], [2,1],         [3,1],         [4,1]],          625).
instance(mc22b,5, [1000, 250, 500,1000], [1,5,3],   [[3,1,[2,4],1], [4,1],         [2,1],         [3,1]],          400).
instance(ma23b,5, [1000, 250, 500],      [6,5],     [[2,1,3,1],     [3,1],         [2,1]],                         249).
instance(mb23b,5, [1000, 500,1000],      [6,7],     [[2,1,3,1],     [[2,3],1],     [2,1,3,1]],                     344).
instance(mc23b,5, [1000, 250,1000],      [8,5],     [[2,1,3,1],     [3,1],         [2,1]],                         250).
instance(md23b,5, [ 500, 500, 500, 500], [2,1,4],   [[2,1,3,1,4,1], [[3,4],1],     [4,1,[2,3],1], [[2,4],1]],      495).
instance(ma24b,5, [1000, 250,1000],      [2,1],     [[2,1,3,1],     [3,1],         [2,1]],                        1000).
instance(mb24b,5, [ 500,1000,1000],      [2,2],     [[2,1,3,1],     [2,1,3,1],     [2,1,3,1]],                    1250).
\end{SaveVerbatim}
\resizebox{1\textwidth}{!}{\BUseVerbatim{VerbInstancesA}}%

\noindent\begin{SaveVerbatim}{VerbInstancesB}
instance(ma6c, 5, [1000, 500,1000],      [3,1],     [[[2,3],1],     [3,1],         [2,1]],                         624).
instance(ma7c, 5, [1000, 250, 500],      [12,3],    [[[2,3],1],     [3,1],         [2,1]],                         116).
instance(ma8c, 5, [1000, 250, 500],      [2,8],     [[[2,3],1],     [3,1],         [2,1]],                         154).
instance(ma12c,5, [1000, 500,1000],      [3,2],     [[2,1,3,1],     [3,1],         [2,1]],                         666).
instance(mb12c,5, [1000, 250,1000,1000], [2,1,2],   [[2,1,3,1,4,1], [3,1],         [2,1],         [4,1]],         1000).
instance(ma13c,5, [ 250, 250, 250],      [2,1],     [[[2,3],1],     [2,1],         [3,1]],                         247).
instance(mb13c,5, [1000, 250,1000,1000], [1,4,6],   [[3,1,[2,4],1], [[3,4],1],     [4,1,[2,3],1], [3,1,[2,4],1]],  534).
instance(ma14c,5, [1000, 500, 500],      [8,8],     [[2,1,3,1],     [2,1],         [3,1]],                         187).
instance(mb14c,5, [1000, 250, 500,1000], [4,4,6],   [[2,1,3,1,4,1], [2,1],         [3,1],         [4,1]],          312).
instance(mb22c,5, [1000, 500,1000],      [7,6],     [[2,1,3,1],     [[2,3],1],     [2,1,3,1]],                     344).
instance(mc22c,5, [ 500, 250, 250],      [2,2],     [[2,1,3,1],     [2,1],         [3,1]],                         375).
instance(mc22d,5, [1000, 250, 500, 500], [3,3,3],   [[2,1,3,1,4,1], [2,1],         [3,1],         [4,1]],          416).
instance(mb23c,5, [1000, 500, 500],      [6,6],     [[2,1,3,1],     [2,1],         [3,1]],                         249).
instance(mc23c,5, [1000, 250, 500],      [7,1],     [[[2,3],1],     [3,1],         [2,1]],                         213).
instance(md23c,5, [1000, 250, 500, 500], [3,3,3],   [[2,1,3,1,4,1], [3,1],         [2,1],         [4,1]],          416).
instance(ma7d, 5, [1000, 250,1000],      [6,9],     [[2,1,3,1],     [2,1],         [3,1]],                         207).
instance(ma8d, 5, [ 500, 500, 500],      [6,2],     [[2,1,3,1],     [[2,3],1],     [[2,3],1]],                     245).
instance(ma13d,5, [1000, 500, 500],      [6,5],     [[2,1,3,1],     [2,1],         [3,1]],                         249).
instance(mb13d,5, [1000, 250, 500, 500], [3,4,4],   [[2,1,3,1,4,1], [2,1],         [3,1],         [4,1]],          375).
instance(ma14d,5, [1000, 250, 500],      [6,6],     [[2,1,3,1],     [2,1],         [3,1]],                         207).
instance(mb14d,5, [1000, 250, 500, 500], [4,3,1],   [[2,1,[3,4],1], [4,1],         [2,1],         [3,1]],          375).
instance(mb21d,5, [1000, 500, 500,1000], [2,1,2],   [[2,1,3,1,4,1], [[2,3],1],     [[2,3],1],     [4,1]],          994).
instance(mc21d,5, [1000, 250, 500, 500], [2,1,1],   [[2,1,3,1,4,1], [3,1],         [[2,4],1],     [[2,4],1]],      994).
instance(ma22d,5, [1000, 250,1000],      [6,6],     [[2,1,3,1],     [[2,3],1],     [2,1,3,1]],                     352).
instance(mb22d,5, [1000, 250, 500],      [4,4],     [[2,1,3,1],     [3,1],         [2,1]],                         312).
instance(md22d,5, [ 500, 250, 250, 500], [2,1,2],   [[2,1,3,1,4,1], [[2,3],1],     [[2,3],1],     [4,1]],          494).
instance(ma23d,5, [ 500,1000,1000],      [7,7],     [[2,1,3,1],     [2,1,3,1],     [2,1,3,1]],                     355).
instance(mb23d,5, [1000, 250, 250,1000], [2,6,2],   [[3,1,[2,4],1], [2,1],         [4,1],         [3,1]],          332).
instance(mc23d,5, [1000, 250, 500,1000], [1,6,4],   [[3,1,[2,4],1], [2,1],         [4,1],         [3,1]],          332).
instance(ma25d,5, [1000, 500, 500,1000], [2,1,1],   [[2,1,[3,4],1], [3,1],         [4,1],         [2,1]],         1000).
instance(mb21c,5, [1000, 500, 500, 500], [2,1,1],   [[2,1,3,1,4,1], [[2,3],1],     [2,1,[3,4],1], [[2,4],1]],     1244).
instance(mc21c,5, [1000, 500, 500, 500], [1,1,2],   [[2,1,3,1,4,1], [[2,4],1],     [4,1,[2,3],1], [[3,4],1]],     1244).
instance(ma25b,5, [1000,1000,1000,1000], [2,2,1],   [[2,1,3,1,4,1], [2,1,[3,4],1], [3,1,[2,4],1], [2,1,3,1,4,1]], 1997).
\end{SaveVerbatim}
\resizebox{1\textwidth}{!}{\BUseVerbatim{VerbInstancesB}}%

\subsection{Optimal Solutions Founds}\label{sec:solutions}

For each instance of Section~\ref{sec:instances} we provide:
\begin{itemize}[label={\scriptsize\textbullet}]
\item The instance name,
\item The time in milliseconds for finding the optimal solution and proving optimality, or for proving that no solution exists. Experiments were done with SICStus Prolog 4.8.0 on an M2 Ultra using one single core.
\item The value of the optimal solution, i.e. the sum of the variables $p_1, p_2, \dots, p_n$ found by the MIP model of Section~\ref{sec:capacity_constraints}. Value $0$ means that the instance was infeasible.
\item The time spent in each stage, i.e. the value of the integer variables $p_1, p_2, \dots, p_n$.
This time was obtained using the keyword \texttt{runtime} of the \texttt{statistics/2} predicate.
\item The sequence of locations traversed by each container, i.e.~the matrix $V$.
\end{itemize}
For example, for the instance~{\footnotesize\texttt{a4}} corresponding to the solution depicted in Part~(B) of Figure~\ref{fig:example_capacity} we have the following information:
{\scriptsize
\begin{verbatim}
     Instance = a4, Time = 11, Opt = 64, Pi = [6,6,6,6,20,6,14],
     V = [[2,1,3,1,4,4,1],[1,2,2,3,3,1,4],[4,4,1,2,2,2,2],[3,3,4,4,1,3,3]]
\end{verbatim}
}

\noindent As the runtime reported by \texttt{statistics/2} may not be very accurate when dealing with few milliseconds, we also provide the total time for solving all the instances using one single core.
The cumulative average resolution time for all instances was 5.2 seconds over 10 runs, i.e.~the sum of the resolution times of all instances. These 5.2 seconds correspond to the CPU time spent in user-mode.

{\scriptsize
\begin{verbatim}
Instance = a1, Time = 11, Opt = 588, Pi = [288,6,288,6],
V = [[2,1,3,1],[1,2,2,2],[3,3,1,3]]
Instance = a2, Time = 6, Opt = 112, Pi = [49,6,44,6,7],
V = [[2,1,3,3,1],[1,2,2,1,3],[3,3,1,2,2]]
Instance = a3, Time = 6, Opt = 112, Pi = [50,6,6,44,6],
V = [[2,1,3,3,1],[1,3,1,2,2],[3,2,2,1,3]]
Instance = a12, Time = 11, Opt = 112, Pi = [49,6,44,6,7],
V = [[2,1,3,3,1],[1,2,2,1,3],[3,3,1,2,2]]
Instance = a4, Time = 11, Opt = 64, Pi = [6,6,6,6,20,6,14],
V = [[2,1,3,1,4,4,1],[1,2,2,3,3,1,4],[4,4,1,2,2,2,2],[3,3,4,4,1,3,3]]
Instance = a5, Time = 10, Opt = 52, Pi = [6,6,8,6,6,13,7],
V = [[2,1,3,3,1,4,1],[1,2,2,2,3,1,4],[4,4,4,1,2,2,2],[3,3,1,4,4,3,3]]
Instance = a6, Time = 7, Opt = 0, Pi = [],
V = []
Instance = a7, Time = 10, Opt = 61, Pi = [17,6,6,9,6,11,6],
V = [[2,2,1,3,1,4,1],[4,1,2,2,3,1,4],[1,4,4,4,2,2,2],[3,3,3,1,4,3,3]]
Instance = a8, Time = 9, Opt = 70, Pi = [6,14,6,6,6,26,6],
V = [[2,2,1,3,1,4,1],[1,4,4,1,3,3,2],[4,1,2,2,2,2,4],[3,3,3,4,4,1,3]]
Instance = a9, Time = 7, Opt = 0, Pi = [],
V = []
Instance = a10, Time = 9, Opt = 64, Pi = [6,11,6,6,20,6,9],
V = [[2,1,3,1,4,4,1],[1,4,1,3,3,2,2],[4,2,2,2,2,1,4],[3,3,4,4,1,3,3]]
Instance = a11, Time = 8, Opt = 64, Pi = [6,6,20,6,6,11,9],
V = [[2,1,3,3,1,4,1],[1,4,4,1,3,2,2],[4,2,2,2,2,1,4],[3,3,1,4,4,3,3]]
Instance = a13, Time = 66, Opt = 70, Pi = [9,11,6,6,6,26,6],
V = [[2,2,1,3,1,4,1],[1,4,4,1,3,3,2],[4,1,2,2,2,2,4],[3,3,3,4,4,1,3]]
Instance = ia6, Time = 4, Opt = 562, Pi = [63,62,437],
V = [[2,3,1],[1,2,2],[3,1,3]]
Instance = ia7, Time = 1, Opt = 249, Pi = [83,112,49,5],
V = [[2,1,3,1],[1,2,2,2],[3,3,1,3]]
Instance = ia8, Time = 2, Opt = 156, Pi = [31,103,5,17],
V = [[2,1,3,1],[3,3,1,3],[1,2,2,2]]
Instance = ib12, Time = 308, Opt = 666, Pi = [201,5,125,82,248,5],
V = [[2,1,3,1,4,1],[1,2,2,2,2,2],[4,4,4,3,1,4],[3,3,1,4,3,3]]
Instance = ia13, Time = 1, Opt = 100, Pi = [50,5,5,40],
V = [[2,1,3,1],[1,2,2,2],[3,3,1,3]]
Instance = ib13, Time = 24, Opt = 0, Pi = [],
V = []
Instance = ia14, Time = 1, Opt = 100, Pi = [50,5,5,40],
V = [[2,1,3,1],[1,2,2,2],[3,3,1,3]]
Instance = ib14, Time = 24, Opt = 271, Pi = [6,116,100,5,35,9],
V = [[2,1,3,1,4,1],[1,2,2,2,2,2],[3,3,4,4,1,3],[4,4,1,3,3,4]]
Instance = ia21, Time = 1, Opt = 250, Pi = [125,5,5,115],
V = [[2,1,3,1],[1,2,2,2],[3,3,1,3]]
Instance = ia22, Time = 2, Opt = 249, Pi = [5,156,83,5],
V = [[2,1,3,1],[3,3,1,3],[1,2,2,2]]
Instance = ic22, Time = 1, Opt = 249, Pi = [80,81,83,5],
V = [[2,1,3,1],[1,2,2,2],[3,3,1,3]]
Instance = id22, Time = 60, Opt = 333, Pi = [5,157,83,83,5],
V = [[2,1,3,4,1],[3,3,1,3,3],[4,4,4,1,4],[1,2,2,2,2]]
Instance = ib23, Time = 25, Opt = 0, Pi = [],
V = []
Instance = ic23, Time = 45, Opt = 227, Pi = [5,5,102,5,105,5],
V = [[2,1,3,1,4,1],[3,3,1,3,3,3],[4,4,4,4,1,2],[1,2,2,2,2,4]]
Instance = ja13, Time = 2, Opt = 102, Pi = [52,5,40,5],
V = [[2,1,3,1],[1,2,2,2],[3,3,1,3]]
Instance = jb13, Time = 24, Opt = 0, Pi = [],
V = []
Instance = jb23, Time = 26, Opt = 0, Pi = [],
V = []
Instance = jc23, Time = 16, Opt = 158, Pi = [74,5,33,5,36,5],
V = [[2,1,3,1,4,1],[3,3,1,3,3,3],[1,2,2,2,2,4],[4,4,4,4,1,2]]
Instance = kb13, Time = 56, Opt = 0, Pi = [],
V = []
Instance = kb14b, Time = 73, Opt = 0, Pi = [],
V = []
Instance = kb23b, Time = 2, Opt = 198, Pi = [115,5,73,5],
V = [[2,1,3,1],[1,2,2,2],[3,3,1,3]]
Instance = ka23c, Time = 1, Opt = 198, Pi = [73,5,115,5],
V = [[2,1,3,1],[3,3,1,3],[1,2,2,2]]
Instance = kb23c, Time = 2, Opt = 215, Pi = [132,5,73,5],
V = [[2,1,3,1],[1,2,2,2],[3,3,1,3]]
Instance = kd23c, Time = 23, Opt = 0, Pi = [],
V = []
Instance = kb21d, Time = 26, Opt = 0, Pi = [],
V = []
Instance = kd22d, Time = 23, Opt = 0, Pi = [],
V = []
Instance = la6, Time = 3, Opt = 562, Pi = [438,62,62],
V = [[2,3,1],[1,2,2],[3,1,3]]
Instance = la7, Time = 2, Opt = 173, Pi = [90,5,73,5],
V = [[2,1,3,1],[1,2,2,2],[3,3,1,3]]
Instance = la8, Time = 1, Opt = 135, Pi = [73,5,52,5],
V = [[2,1,3,1],[3,3,1,3],[1,2,2,2]]
Instance = la12, Time = 12, Opt = 832, Pi = [161,5,328,5,161,172],
V = [[2,2,1,3,3,1],[1,3,3,1,2,2],[3,1,2,2,1,3]]
Instance = lb12, Time = 195, Opt = 482, Pi = [232,5,230,5,5,5],
V = [[2,1,3,1,4,1],[1,2,2,2,2,2],[3,3,1,4,3,3],[4,4,4,3,1,4]]
Instance = la13, Time = 2, Opt = 102, Pi = [40,5,52,5],
V = [[2,1,3,1],[1,2,2,2],[3,3,1,3]]
Instance = lb13, Time = 28, Opt = 208, Pi = [146,5,5,6,5,5,26,5,5],
V = [[2,2,1,3,3,1,4,4,1],[1,4,4,4,4,4,2,2,2],[3,3,3,1,2,2,1,3,3],[4,1,2,2,1,3,3,1,4]]
Instance = la14, Time = 1, Opt = 90, Pi = [40,5,40,5],
V = [[2,1,3,1],[1,2,2,2],[3,3,1,3]]
Instance = lb14, Time = 17, Opt = 213, Pi = [130,5,63,5,5,5],
V = [[2,1,3,1,4,1],[1,2,2,2,2,2],[3,3,4,4,1,3],[4,4,1,3,3,4]]
Instance = la21, Time = 5, Opt = 497, Pi = [5,247,245],
V = [[2,3,1],[1,2,2],[3,1,3]]
Instance = lb21, Time = 1, Opt = 240, Pi = [115,5,115,5],
V = [[2,1,3,1],[1,2,2,2],[3,3,1,3]]
Instance = ld21, Time = 10, Opt = 742, Pi = [243,5,242,5,242,5],
V = [[2,1,3,1,4,1],[1,2,2,2,2,2],[3,3,1,3,3,3],[4,4,4,4,1,4]]
Instance = la22, Time = 2, Opt = 198, Pi = [73,5,115,5],
V = [[2,1,3,1],[3,3,1,3],[1,2,2,2]]
Instance = lb22, Time = 1, Opt = 322, Pi = [156,5,156,5],
V = [[2,1,3,1],[1,2,2,2],[3,3,1,3]]
Instance = lc22, Time = 2, Opt = 281, Pi = [156,5,115,5],
V = [[2,1,3,1],[1,2,2,2],[3,3,1,3]]
Instance = ld22, Time = 4, Opt = 265, Pi = [99,5,15,141,5],
V = [[2,1,3,4,1],[3,3,1,3,3],[4,4,4,1,4],[1,2,2,2,2]]
Instance = la23, Time = 2, Opt = 250, Pi = [115,5,125,5],
V = [[2,1,3,1],[1,2,2,2],[3,3,1,3]]
Instance = lb23, Time = 6, Opt = 203, Pi = [37,5,78,78,5],
V = [[2,1,3,4,1],[3,3,1,3,3],[4,4,4,1,4],[1,2,2,2,2]]
Instance = lc23, Time = 3, Opt = 159, Pi = [34,5,76,5,34,5],
V = [[2,1,3,1,4,1],[3,3,1,3,3,3],[4,4,4,4,1,4],[1,2,2,2,2,2]]
Instance = la24, Time = 5, Opt = 617, Pi = [118,5,117,5,367,5],
V = [[2,1,3,1,4,1],[4,4,4,4,1,4],[1,2,2,2,2,2],[3,3,1,3,3,3]]
Instance = la25, Time = 6, Opt = 617, Pi = [368,5,117,5,117,5],
V = [[2,1,3,1,4,1],[1,2,2,2,2,2],[3,3,1,3,3,3],[4,4,4,4,1,4]]
Instance = la26, Time = 1, Opt = 490, Pi = [240,5,240,5],
V = [[2,1,3,1],[1,2,2,2],[3,3,1,3]]
Instance = lb26, Time = 4, Opt = 375, Pi = [175,125,75],
V = [[2,3,1],[3,1,3],[1,2,2]]
Instance = la30, Time = 1, Opt = 490, Pi = [240,5,240,5],
V = [[2,1,3,1],[1,2,2,2],[3,3,1,3]]
Instance = lb30, Time = 2, Opt = 656, Pi = [323,5,323,5],
V = [[2,1,3,1],[1,2,2,2],[3,3,1,3]]
Instance = le30, Time = 5, Opt = 367, Pi = [118,5,117,5,117,5],
V = [[2,1,3,1,4,1],[1,2,2,2,2,2],[3,3,1,3,3,3],[4,4,4,4,1,4]]
Instance = ma6b, Time = 2, Opt = 740, Pi = [240,5,490,5],
V = [[2,1,3,1],[3,3,1,3],[1,2,2,2]]
Instance = ma7b, Time = 7, Opt = 285, Pi = [37,185,63],
V = [[2,3,1],[1,2,2],[3,1,3]]
Instance = ma8b, Time = 5, Opt = 251, Pi = [151,5,5,85,5],
V = [[3,1,2,4,1],[2,2,1,2,2],[1,3,3,3,3],[4,4,4,1,4]]
Instance = ma12b, Time = 1, Opt = 322, Pi = [156,5,156,5],
V = [[2,1,3,1],[1,2,2,2],[3,3,1,3]]
Instance = mb12b, Time = 1, Opt = 490, Pi = [240,5,240,5],
V = [[2,1,3,1],[1,2,2,2],[3,3,1,3]]
Instance = mc12b, Time = 2, Opt = 406, Pi = [156,5,240,5],
V = [[2,1,3,1],[3,3,1,3],[1,2,2,2]]
Instance = ma13b, Time = 3, Opt = 163, Pi = [5,80,78],
V = [[2,3,1],[1,2,2],[3,1,3]]
Instance = mb13b, Time = 17, Opt = 108, Pi = [57,5,28,5,8,5],
V = [[2,1,3,1,4,1],[4,4,4,4,1,4],[1,3,2,2,2,2],[3,2,1,3,3,3]]
Instance = ma14b, Time = 3, Opt = 192, Pi = [26,92,74],
V = [[2,3,1],[1,2,2],[3,1,3]]
Instance = mb14b, Time = 75, Opt = 320, Pi = [125,49,6,10,115,5,10],
V = [[2,1,3,1,4,4,1],[4,4,4,4,1,2,2],[3,3,1,2,2,1,4],[1,2,2,3,3,3,3]]
Instance = mb21b, Time = 1, Opt = 573, Pi = [323,5,240,5],
V = [[2,1,3,1],[1,2,2,2],[3,3,1,3]]
Instance = ma22b, Time = 2, Opt = 190, Pi = [90,5,90,5],
V = [[2,1,3,1],[1,2,2,2],[3,3,1,3]]
Instance = mb22b, Time = 3, Opt = 180, Pi = [55,5,55,5,55,5],
V = [[2,1,3,1,4,1],[1,2,2,2,2,2],[3,3,1,3,3,3],[4,4,4,4,1,4]]
Instance = mc22b, Time = 5, Opt = 268, Pi = [68,5,5,185,5],
V = [[3,1,2,4,1],[4,4,4,1,4],[2,2,1,2,2],[1,3,3,3,3]]
Instance = ma23b, Time = 1, Opt = 123, Pi = [40,5,73,5],
V = [[2,1,3,1],[3,3,1,3],[1,2,2,2]]
Instance = mb23b, Time = 10, Opt = 218, Pi = [5,5,137,5,66],
V = [[2,1,3,3,1],[3,3,1,2,2],[1,2,2,1,3]]
Instance = mc23b, Time = 2, Opt = 165, Pi = [40,5,115,5],
V = [[2,1,3,1],[3,3,1,3],[1,2,2,2]]
Instance = md23b, Time = 102, Opt = 356, Pi = [5,8,117,77,10,125,14],
V = [[2,1,3,3,1,4,1],[3,3,1,4,4,3,3],[1,4,4,1,3,2,2],[4,2,2,2,2,1,4]]
Instance = ma24b, Time = 2, Opt = 740, Pi = [240,5,490,5],
V = [[2,1,3,1],[3,3,1,3],[1,2,2,2]]
Instance = mb24b, Time = 15, Opt = 1250, Pi = [245,5,495,5,245,255],
V = [[2,2,1,3,3,1],[1,3,3,1,2,2],[3,1,2,2,1,3]]
Instance = ma6c, Time = 4, Opt = 624, Pi = [291,127,206],
V = [[2,3,1],[3,1,3],[1,2,2]]
Instance = ma7c, Time = 3, Opt = 116, Pi = [75,33,8],
V = [[2,3,1],[3,1,3],[1,2,2]]
Instance = ma8c, Time = 3, Opt = 154, Pi = [5,123,26],
V = [[2,3,1],[3,1,3],[1,2,2]]
Instance = ma12c, Time = 1, Opt = 573, Pi = [240,5,323,5],
V = [[2,1,3,1],[3,3,1,3],[1,2,2,2]]
Instance = mb12c, Time = 6, Opt = 617, Pi = [118,5,367,5,117,5],
V = [[2,1,3,1,4,1],[3,3,1,3,3,3],[1,2,2,2,2,2],[4,4,4,4,1,4]]
Instance = ma13c, Time = 3, Opt = 247, Pi = [122,5,120],
V = [[2,3,1],[1,2,2],[3,1,3]]
Instance = mb13c, Time = 199, Opt = 294, Pi = [5,5,5,5,130,112,5,27],
V = [[3,3,1,2,2,4,4,1],[4,4,4,4,1,1,3,3],[1,2,2,3,3,3,1,4],[2,1,3,1,4,2,2,2]]
Instance = ma14c, Time = 1, Opt = 114, Pi = [52,5,52,5],
V = [[2,1,3,1],[1,2,2,2],[3,3,1,3]]
Instance = mb14c, Time = 3, Opt = 167, Pi = [105,5,42,5,5,5],
V = [[2,1,3,1,4,1],[1,2,2,2,2,2],[3,3,1,3,3,3],[4,4,4,4,1,4]]
Instance = mb22c, Time = 9, Opt = 218, Pi = [66,5,137,5,5],
V = [[2,1,3,3,1],[3,3,1,2,2],[1,2,2,1,3]]
Instance = mc22c, Time = 1, Opt = 240, Pi = [115,5,115,5],
V = [[2,1,3,1],[1,2,2,2],[3,3,1,3]]
Instance = mc22d, Time = 3, Opt = 200, Pi = [117,5,34,5,34,5],
V = [[2,1,3,1,4,1],[1,2,2,2,2,2],[3,3,1,3,3,3],[4,4,4,4,1,4]]
Instance = mb23c, Time = 1, Opt = 156, Pi = [73,5,73,5],
V = [[2,1,3,1],[1,2,2,2],[3,3,1,3]]
Instance = mc23c, Time = 4, Opt = 213, Pi = [142,6,65],
V = [[2,3,1],[3,1,3],[1,2,2]]
Instance = md23c, Time = 2, Opt = 200, Pi = [34,5,117,5,34,5],
V = [[2,1,3,1,4,1],[3,3,1,3,3,3],[1,2,2,2,2,2],[4,4,4,4,1,4]]
Instance = ma7d, Time = 2, Opt = 142, Pi = [101,5,31,5],
V = [[2,1,3,1],[1,2,2,2],[3,3,1,3]]
Instance = ma8d, Time = 13, Opt = 202, Pi = [83,5,60,54],
V = [[2,1,3,1],[1,2,2,3],[3,3,1,2]]
Instance = ma13d, Time = 1, Opt = 173, Pi = [90,5,73,5],
V = [[2,1,3,1],[1,2,2,2],[3,3,1,3]]
Instance = mb13d, Time = 3, Opt = 159, Pi = [76,5,34,5,34,5],
V = [[2,1,3,1,4,1],[1,2,2,2,2,2],[3,3,1,3,3,3],[4,4,4,4,1,4]]
Instance = ma14d, Time = 2, Opt = 114, Pi = [73,5,31,5],
V = [[2,1,3,1],[1,2,2,2],[3,3,1,3]]
Instance = mb14d, Time = 4, Opt = 265, Pi = [141,5,99,15,5],
V = [[2,1,3,4,1],[4,4,4,1,4],[1,2,2,2,2],[3,3,1,3,3]]
Instance = mb21d, Time = 68, Opt = 616, Pi = [364,5,121,5,116,5],
V = [[2,1,3,1,4,1],[1,3,2,2,2,2],[3,2,1,3,3,3],[4,4,4,4,1,4]]
Instance = mc21d, Time = 16, Opt = 482, Pi = [230,5,232,5,5,5],
V = [[2,1,3,1,4,1],[3,3,1,3,3,3],[1,2,2,2,2,4],[4,4,4,4,1,2]]
Instance = ma22d, Time = 10, Opt = 202, Pi = [6,5,161,5,25],
V = [[2,1,3,3,1],[3,3,1,2,2],[1,2,2,1,3]]
Instance = mb22d, Time = 1, Opt = 177, Pi = [52,5,115,5],
V = [[2,1,3,1],[3,3,1,3],[1,2,2,2]]
Instance = md22d, Time = 30, Opt = 303, Pi = [177,5,58,5,53,5],
V = [[2,1,3,1,4,1],[1,3,2,2,2,2],[3,2,1,3,3,3],[4,4,4,4,1,4]]
Instance = ma23d, Time = 8, Opt = 355, Pi = [66,5,137,5,66,76],
V = [[2,2,1,3,3,1],[1,3,3,1,2,2],[3,1,2,2,1,3]]
Instance = mb23d, Time = 6, Opt = 203, Pi = [37,5,78,78,5],
V = [[3,1,2,4,1],[2,2,1,2,2],[4,4,4,1,4],[1,3,3,3,3]]
Instance = mc23d, Time = 4, Opt = 265, Pi = [99,5,15,141,5],
V = [[3,1,2,4,1],[2,2,1,2,2],[4,4,4,1,4],[1,3,3,3,3]]
Instance = ma25d, Time = 12, Opt = 745, Pi = [245,5,245,245,5],
V = [[2,1,3,4,1],[3,3,1,3,3],[4,4,4,1,4],[1,2,2,2,2]]
Instance = mb21c, Time = 82, Opt = 735, Pi = [465,5,5,5,245,5,5],
V = [[2,1,3,1,4,4,1],[1,2,2,3,3,3,3],[3,3,1,2,2,1,4],[4,4,4,4,1,2,2]]
Instance = mc21c, Time = 137, Opt = 735, Pi = [5,245,5,71,5,399,5],
V = [[2,2,1,3,1,4,1],[4,1,2,2,2,2,4],[1,4,4,1,3,3,2],[3,3,3,4,4,1,3]]
Instance = ma25b, Time = 2466, Opt = 1747, Pi = [478,21,226,22,478,5,17,478,17,5],
V = [[2,2,1,3,3,1,4,4,4,1],[1,3,3,4,4,4,1,2,2,2],[4,4,2,2,1,3,3,3,1,4],[3,1,4,1,2,2,2,1,3,3]]
\end{verbatim}
}

\subsection{Instructions to Replicate the Experiments}\label{sec:code}

To replicate the experiments, we provide four ancillary code files that were used, namely:
\begin{itemize}[label={\scriptsize\textbullet}]
\item \texttt{instances.pl} is the set of instances we use.
\item \texttt{stat.pl} is the program we use for generating the statistics reported in Table~\ref{table:results} of Section~\ref{sec:cyclic_disjunctive_constraints}.
\item \texttt{solver.pl} is the program we use to find the optimal solutions for the different instances.
\item \texttt{dfa.pl} is a library of operations on automata, to which we add the synchronised product of finite automata wrt a constraint, see lines 974--1266.
\end{itemize}

To rerun the program which provides the results reported in Table~\ref{table:results}, use the following command from a terminal:

{\scriptsize
\begin{verbatim}
Mac:final nicolasbeldiceanu$ time /Applications/sicstus4.8.0_arm/bin/sicstus --noinfo -DGLOBALSTKSIZE=200
SICStus 4.8.0 (arm64-darwin-20.1.0): Sun Dec  4 14:17:07 CET 2022
Licensed to NicolasBeldiceanuComplimentary
| ?- [stat].
yes
| ?- top.
Nb instance with 3 containers is 63
Nb product with 3 containers is 82
Nb product with 0 solutions = 2
In  states:   min = 196 max = 1805 mean = 437.7317073170732 std dev = 435.51892369572755
Out states:   min = 0 max = 13 mean = 5.646341463414634 std dev = 2.6336099273158116
Out alphabet: min = 0 max = 6 mean = 3.475609756097561 std dev = 1.2804878048780475
Nb instance with 4 containers is 55
Nb product with 4 containers is 179
Nb product with 0 solutions = 81
In  states:   min = 2058 max = 229593 mean = 29920.31843575419 std dev = 37359.361146701376
Out states:   min = 0 max = 61 mean = 6.418994413407821 std dev = 9.034253691972463
Out alphabet: min = 0 max = 24 mean = 4.167597765363128 std dev = 4.675487049011022
yes
| ?- halt.

real	0m20.163s
user	0m1.768s
sys	0m0.027s
\end{verbatim}
}

To rerun the solver which search for optimal solutions for all instances, use the following command from a terminal:

{\scriptsize
\begin{verbatim}
Mac:final nicolasbeldiceanu$ time /Applications/sicstus4.8.0_arm/bin/sicstus --noinfo -DGLOBALSTKSIZE=200
SICStus 4.8.0 (arm64-darwin-20.1.0): Sun Dec  4 14:17:07 CET 2022
Licensed to NicolasBeldiceanuComplimentary
| ?- [solver].
yes
| ?- top(_).
Instance = a1, Time = 5, Opt = 588, Pi = [288,6,288,6],
V = [[2,1,3,1],[1,2,2,2],[3,3,1,3]]
..........................................................
Instance = ma25b, Time = 2475, Opt = 1747, Pi = [478,21,226,22,478,5,17,478,17,5],
V = [[2,2,1,3,3,1,4,4,4,1],[1,3,3,4,4,4,1,2,2,2],[4,4,2,2,1,3,3,3,1,4],[3,1,4,1,2,2,2,1,3,3]]
yes
| ?- halt.

real	0m19.849s
user	0m5.188s
sys	0m0.058s
\end{verbatim}
}

\end{document}